\documentclass[journal]{IEEEtran}
\IEEEoverridecommandlockouts
\usepackage{cite}
\usepackage{amsmath,amssymb,amsfonts}

\usepackage{algorithmic}
\usepackage{algorithm}
\usepackage{graphicx}
\usepackage{textcomp}
\usepackage{multirow}
\usepackage{booktabs}
\usepackage{color}
\usepackage[T1]{fontenc}
\usepackage[table,xcdraw]{xcolor}
\def\BibTeX{{\rm B\kern-.05em{\sc i\kern-.025em b}\kern-.08em
T\kern-.1667em\lower.7ex\hbox{E}\kern-.125emX}}
\begin{document}

\title{Data-Driven Dimension Reduction for Industrial Load Modeling Using Inverse Optimization
}

\author{Ruike Lyu, Hongye Guo$^*$, Goran Strbac, Chongqing Kang,~\IEEEmembership{Fellow,~IEEE}

\thanks{
R. Lyu, H. Guo, and C. Kang are with the State Key Lab of Power System Operation and Control, Dept. of Electrical Eng., Tsinghua Univ., Beijing, China.
G. Strbac is with the Dept. of Electrical and Electronic Eng., Imperial College London, London, U.K.
This work was supported in part by the National Natural Science Foundation of China (NSFC) under Grant 52107102, and in part by the NSFC under Grant 52321004.
}

}

\maketitle

\begin{abstract}
The intricate mixed-integer constraints in industrial load models not only pose challenges for their direct integration into economic dispatch or market clearing processes but also render current analytical dimension-reduction methods ineffective. We propose a novel data-driven dimension-reduction approach for industrial load modeling, which uses the optimal energy usage data from industrial loads to train a dimension-reduced model that best fits the original constraints. Our approach, implemented by the adjustable load fleet model, outperformed analytical methods across three industrial load datasets.
\end{abstract}

\vspace{-3ex}

\section{Introduction}

To harness industrial users' flexibility through grid-load interactions while mitigating adverse effects, it is crucial to mathematically model the constraints of their production processes~\cite{lu_data-driven_2021}. The complex physics of industrial processes poses a key challenge, requiring integer variables or nonlinear functions for representation~\cite{zhang_cost-effective_2017}. Therefore, precise industrial load models encounter obstacles when integrated into economic dispatch or market clearing. To address this issue, we explore dimension reduction techniques to approximate the intricate mixed-integer constraints of industrial loads with simplified lower-dimensional constraints.

Due to the complexity of temporally coupled constraints, exact dimension reduction through projection becomes computationally intractable~\cite{tan_non-iterative_2023}. Therefore, the common approach is to obtain a computationally feasible result using inner or outer approximation methods~\cite{9979729}. Currently, most approximation-based dimension reduction approaches are analytical methods, which first specify a template geometry (e.g., boxes~\cite{chen_aggregate_2020}, ellipses~\cite{chen_leveraging_2021}, or zonotopes~\cite{muller_aggregation_2019}) to approximate the original constraints, then determine the constraint parameters through mathematical derivation based on certain assumptions about the original constraints. While these analytical methods with template-based geometries enable efficient computation of Minkowski sums, they face two key limitations~\cite{lyu_approximating_2024}: 1) the pre-specified geometric templates lack adaptability across various industrial loads with different constraint structures; 2) they require convexity in the original constraints, making them incompatible with industrial load models with integer variables.

\begin{figure}[!t]
  \centering
  \includegraphics[width=3.5in]{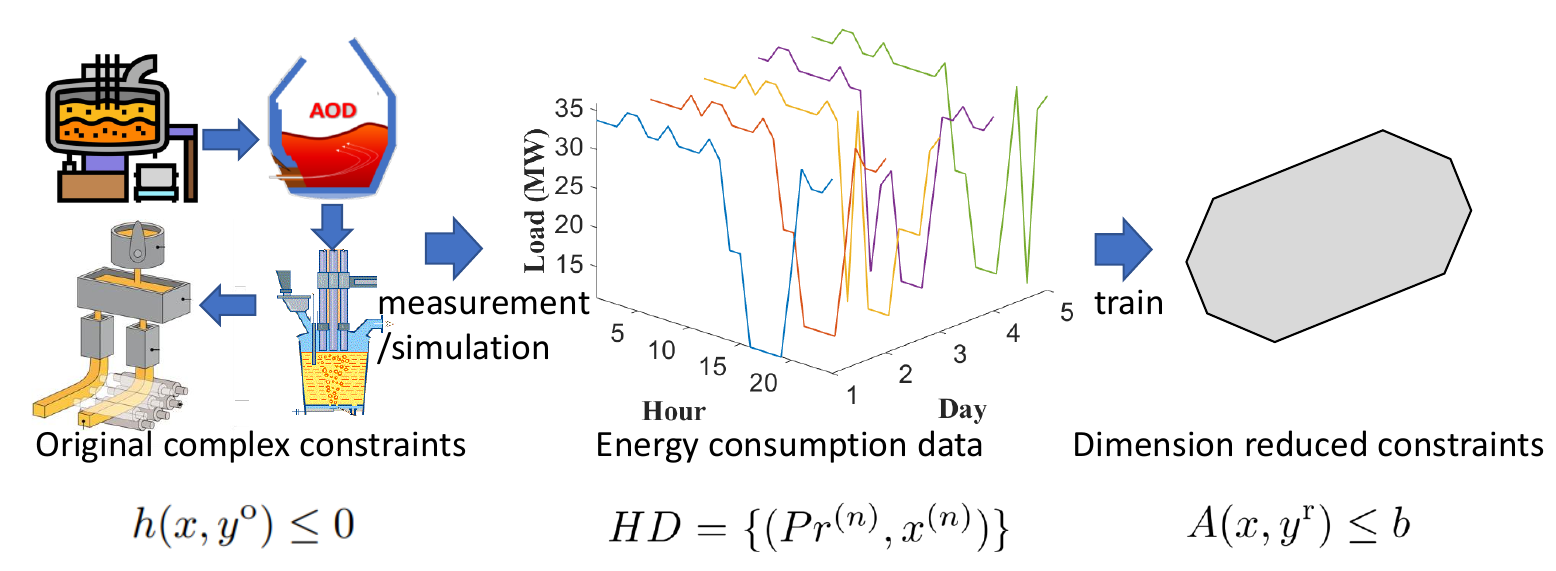}
  \vspace{-5ex}
\caption{Proposed data-driven dimension reduction framework for industrial load modeling, illustrated with a steel plant example. $h$: high-dimensional constraints; $HD$: historical dataset; $A$: low-dimensional constraints.}
  \label{fig_framework}
\end{figure}

In this paper, we propose a data-driven dimension reduction (D3R) framework that leverages optimal energy consumption data of industrial loads derived from the original constraints to train a reduced-constraint model through inverse optimization (Fig.~\ref{fig_framework}). D3R exhibits enhanced adaptability and is compatible with constraints involving integer variables.

\vspace{-2ex}
\section{Problem Description}\label{sec_problem_description}

Without loss of generality, our objective is to replace the original constraints of a given industrial load (original constraints, 
OCs
), denoted as \(h(x, y^{\rm o}) \leq 0\), with a set of dimension-reduced linear constraints (reduced constraints, 
RCs
) \(A(x, y^{\rm r}) \leq b\). Here, \(h(\cdot)\) is the original constraint function which typically involves complex mixed-integer constraints from Resource-Task Network (RTN)~\cite{zhang_cost-effective_2017} or State-Task Network (STN) models~\cite{golmohamadi_robust_2020, lu_data-driven_2021} of industrial processes,
\(x \in \mathbb{R}^T\) represents the hourly energy consumption (continuous), $T$ is the number of time intervals, \(y^{\rm o}\) represents other variables of the OCs (including both continuous variables for states/resources and integer variables for operating modes/task scheduling), \(y^{\rm r}\) represents other variables of the RCs (continuous), distinguished between the superscripts o (original) and r (reduced), respectively; \(A\) and \(b\) are the parameter matrix of and the right-hand side of the RCs, respectively.

If we require strict equivalence between the RCs and OCs, then this involves projecting the variable space associated with the OCs onto the RCs, which means the following:
\begin{equation}\label{model_strict_projection}
  \forall x: \exists y^{\rm r}, A(x, y^{\rm r}) \leq b \Leftrightarrow \exists y^{\rm o}, h(x, y^{\rm o}) \leq 0.
\end{equation}
However, strict projection may be computationally infeasible in cases where \(h(\cdot)\) has high dimensionality, $y^{\rm o}$ contains integer variables, or $T$ is large. Meanwhile, considering the current market mechanisms for demand-side resources, strict projection may not be necessary. In electricity markets, demand-side resources typically provide services such as demand response, reserves and frequency regulation, where performance requirements are more flexible compared to generation resources. For instance, a 10\% to 20\% tolerance is generally acceptable, which motivates us to pursue an approximation-based approach rather than strict equivalence or guaranteeing feasibility that may be overly conservative.

Considering these practical factors, we relax Equ. (\ref{model_strict_projection}) and introduce a loss function \(J(h(\cdot), A, b)\) to quantify the error of the RCs relative to the OCs. Therefore, our objective transforms into determining appropriate \(A\) and \(b\) values to minimize the approximation error $J(h(\cdot), A, b)$.

Specifically, we do not derive \(A\) and \(b\) analytically on the basis of the form and parameters of the constraint \(h(\cdot)\). Instead, we observe the historical energy usage of industrial users and parameterize (train) \(A\) and \(b\) in a data-driven manner. In doing so, we assume access to historical data of industrial users \(HD = \{(Pr^{(n)}, x^{(n)})\}\), where \(Pr^{(n)}\) and \(x^{(n)}\) represent the hourly electricity price and the user's hourly electricity consumption on day \(n\), respectively, $n = 1, ..., N$. We also assume that \(x^{(n)}\) are optimized on the basis of the OCs to minimize the energy cost $Pr^{(n)\intercal} x^{(n)}$.
Note that \(HD\) can be realistic or simulated historical data.
In practice, historical energy consumption data from industrial users may not originate from optimized decisions. For instance, factories might employ heuristic strategies rather than complex optimization models to schedule their production and energy usage. In such cases, the D3R framework remains applicable but requires an additional step: first, simulating optimal energy usage data under varying electricity prices using the original constraints (OCs), then using these simulated data to train the reduced constraints (RCs).

Under the above assumptions, the approximation error can be expressed as the distance between the optimal energy consumption under historical electricity prices given by the OCs and RCs, and the training problem is formulated as follows:
\begin{subequations}\label{model_inverse_fit}
  \begin{equation}\label{model_inverse_fit_lossFunction}
    \underset{A, b}{\rm min.} J = \frac{1}{N} \sum_{n = 1}^{N} ||x^{(n)} - x_n||^2
  \end{equation}\vspace{-2ex}
  \begin{equation}\label{model_optimality}
    {\rm s.t.} \quad x_n = {\rm argmin.}\{Pr^{(n)\intercal} x_n \ : \ A(x_n, y^{\rm r}) \leq b\}, \ \forall n
  \end{equation}
\end{subequations}
where \(x_n\) is the variable for the optimization problem regarding day $n$ based on the RCs to distinguish it from \(x^{(n)}\) in the dataset. (\ref{model_inverse_fit}) is a data-driven inverse optimization problem~\cite{lyu_approximating_2024}, which will be further elaborated in Section~\ref{sec_method}.

\vspace{-2ex}
\section{Methodology}\label{sec_method}

In principle, D3R allows for the exploration of suitable parameter matrix sizes and element values within the space \(A/b\). However, the high-dimensional search space and the high-dimensional nonlinear constraints of the optimality conditions in (\ref{model_optimality}) may lead to unaffordable computational costs. To address this challenge, our approach is twofold: first, we selected an appropriate form of the parameter matrix on the basis of the energy consumption characteristics of industrial users (Section~\ref{sec_alf}), and second, we devised a solution algorithm on the basis of zeroth-order stochastic gradient descent (ZOSGD)~\cite{liu_primer_2020} 
to iteratively solve the inverse optimization problem, with a transformation of the optimality conditions to reduce the computational complexity at each iteration (Section~\ref{sec_training}).

\vspace{-2ex}
\subsection{ALF-based Reduced Constraints}\label{sec_alf}

We chose an adjustable load fleet (ALF) composed of multiple adjustable loads (ALs) as the form of the RC to be trained. The motivation for choosing ALF is that it can directly model the actual composition and behavior of industrial loads without requiring baseline load definitions or assuming bidirectional energy exchange capabilities as in virtual battery models~\cite{9979729}. The ALF model captures three key physical characteristics of industrial loads through linear constraints:

1) Equipment Composition: Multiple pieces of equipment in industrial facilities are represented by parallel ALs, where the total energy consumption is the sum of individual device consumption.

2) Equipment Power Rating: Each device's power consumption is bounded by its rated capacity.

3) Production Goals: Daily energy consumption limits reflect production targets, as they are typically proportional to energy usage through product energy intensity factors.

The RCs in the form of an ALF are then:
\begin{subequations}\label{model_alf}\vspace{-2ex}
  \begin{equation}\label{model_energy_powerLimit}
    x_{t} = \sum_{i=1}^{I} p_{t, i} \Delta t, \quad \underline{P}_{i} \le p_{t, i} \le \overline{P}_{i} \ : \ \underline{\mu}^{\rm P}_{t, i}, \overline{\mu}^{\rm P}_{t, i}, \ \forall  t
  \end{equation}\vspace{-2ex}
  \begin{equation}\label{model_energy_energyLimit}
    \underline{E}_{i} \le \sum_{t=1}^{T} p_{t, i} \Delta t \le \overline{E}_{i} \ : \ \underline{\mu}^{\rm E}_{i}, \overline{\mu}^{\rm E}_{i}
  \end{equation}
\end{subequations}
where the subscript $n$ is omitted, $I$ is the number of ALs as a hyperparameter that balances model complexity and approximation accuracy. A larger $I$ allows RCs to better approximate OCs but increases computational costs, while limited training data may constrain the choice of $I$ to avoid overfitting. $x_{t}$ is the net energy consumption at time \(t\), which is the summation of \(p_{t, i} \Delta t\), representing the average consumption power of AL \(i\) at time \(t\) multiplied by the length of time interval $\Delta t$. \(\theta = \{\underline{P}_{i}, \overline{P}_{i}, \underline{E}_{i}, \overline{E}_{i}\}\) include the power and energy limits of AL \(i\), respectively, which are the parameters to be fitted. The Lagrange multipliers \(\underline{\mu}^{\rm P}_{t, i}, \overline{\mu}^{\rm P}_{t, i}, \underline{\mu}^{\rm E}_{i}, \overline{\mu}^{\rm E}_{i}\) are the dual variables associated with the power and energy limits, respectively. The above linear constraints can be easily transformed into the form of \(A(x, y^{\rm r}) \leq b\), where \(x = [x_t]\) and \(y^{\rm r} = [p_{t, i}]\).

While this simplified model may overlook some temporal coupling between production processes and relax integer variables, it effectively balances computational complexity and approximation accuracy. Moreover, the D3R framework is not limited to the ALF model - other reduced constraint forms can be chosen based on specific needs while maintaining the framework's core methodology.

\vspace{-2ex}
\subsection{Training of the ALF model}\label{sec_training}

Substituting the form of the ALF, the optimality conditions (Karush-Kuhn-Tucker conditions) represented by Equation (\ref{model_optimality}) for day \(n\) include primal feasibility (\ref{model_alf}), stationarity condition derived from the Lagrangian function (\ref{model_stationarity}), dual feasibility (\ref{model_dual_feasibility}), and complementary slackness (\ref{model_complementary_slackness1}, \ref{model_complementary_slackness2}):
\begin{subequations}\label{model_optimality_conditions}
  \begin{equation}\label{model_stationarity}
    Pr^{\rm E}_{t} - \underline{\mu}^{\rm P}_{t, i} + \overline{\mu}^{\rm P}_{t, i} - \underline{\mu}^{\rm E}_{i} + \overline{\mu}^{\rm E}_{i} = 0, \ \forall t, \forall i
  \end{equation}\vspace{-3.8ex}
  \begin{equation}\label{model_dual_feasibility}
    (\underline{\mu}^{\rm P}_{t, i}, \overline{\mu}^{\rm P}_{t, i}, \underline{\mu}^{\rm E}_{i}, \overline{\mu}^{\rm E}_{i}) \ge 0, \ \forall t, \forall i
  \end{equation}\vspace{-3.8ex}
\begin{equation}\label{model_complementary_slackness1}
    \underline{\mu}^{\rm P}_{t, i} (\underline{P}_{i} - p_{t, i}) = 0, \quad \overline{\mu}^{\rm P}_{t, i} (p_{t, i} - \overline{P}_{i}) = 0, \quad \forall t, i
  \end{equation}\vspace{-3.8ex}
  \begin{equation}\label{model_complementary_slackness2}
    \underline{\mu}^{\rm E}_{i} (\underline{E}_{i} - \sum_{t=1}^{T} p_{t, i}) = 0, \quad \overline{\mu}^{\rm E}_{i} (\sum_{t=1}^{T} p_{t, i} - \overline{E}_{i}) = 0, \quad \forall i
  \end{equation}
\end{subequations}
The complementary slackness (\ref{model_complementary_slackness1}, \ref{model_complementary_slackness2}) conditions involve bilinear constraints, which can be transformed via the Fortuny-Amat transformation~\cite{fortuny-amat_representation_1981} into a more easily solvable form. For example, the first term can be replaced by the following:
\begin{equation}\label{model_complementary_slackness3}
  \underline{\mu}^{\rm P}_{t, i} \le M(1 - z_{t, i}), \quad p_{t, i} - \underline{P}_{i} \le M z_{t, i}, \quad \forall t, i
\end{equation}
where $z_{t, i}$ is a binary variable and $M$ is a large positive number. The transformed conditions are used in inverse optimization (\ref{model_inverse_fit}), forming a mixed-integer linear programming (MILP) problem that can be solved by commercial solvers.

Since the number of integer variables scales with dataset size \(N\), direct solving becomes computationally intractable. We employ ZOSGD~\cite{liu_primer_2020} for iterative solving, which is suitable because it: 1) avoids explicit gradient computation, 2) reduces computational burden through batch processing, and 3) maintains tractability via iterative optimization. In each iteration, certain days of data are randomly selected and parameters are updated using estimated gradients:

1: Initialize $j = 0$ and $\theta^{(0)} = \{\underline{P}_{i}, \overline{P}_{i}, \underline{E}_{i}, \overline{E}_{i}\} = [0]$.

2: Randomly select $B$ days from the dataset and solve the batch problem:
$\underset{\theta}{\rm min.} J = \frac{1}{B} \sum_{n = 1}^{B} ||x^{(n)} - x_n||^2 \ {\rm s.t.} (\ref{model_alf})-(\ref{model_optimality_conditions})$ to obtain $\theta^{*}$.

3: Update $\theta^{(j+1)} = (1 - \alpha)\theta^{(j)} + \alpha \theta^{*}$; if $\theta^{(j+1)}$ converges, break; otherwise, set $j = j + 1$ and go to step 2.

where $\alpha$ is the learning rate, which can be adaptively adjusted. The ZOSGD algorithm can be easily implemented by using commercial solvers, and the computational complexity is independent of the number of days in the dataset.

\vspace{-1ex}
\section{Numerical Results}\label{sec_numerical}

\paragraph{Dataset and Task Description} We tested the D3R framework on three datasets from a cement plant~\cite{golmohamadi_robust_2020}, a steel powder plant~\cite{lu_data-driven_2021}, and a steelmaking plant~\cite{zhang_cost-effective_2017}. The first two were originally modeled via the STN, whereas the steelmaking plant was modeled via the RTN. Both are high-dimensional models with integer variables and were used to simulate the optimal energy consumption results for training the ALF.

We used historical hourly electricity market prices from PJM in July 2022 for the simulations. The data from July 1st to 21st contained the training set, whereas the data from July 22nd to 31st were used to test the accuracy of the reduced constraints. The detailed settings and codes can be found in ~\cite{rick10119_data-driven_2024}. We employed Gurobi (V11.0.0) with YALMIP
in MATLAB to solve the optimization problems on a workstation equipped with an Intel Core i9-10900X CPU (3.7 GHz) and 128 GB of RAM.

For complex industrial load models, there are currently no mature methods for dimension reduction. Therefore, we compared the accuracy of D3R with that of:

1) the simple AL model (SAL), which assigns values to \(\theta\) based on the maximum and minimum values of the corresponding quantities in the historical load curves. For instance, the maximum power value from historical data is directly taken as the \(P^{\rm max} (\overline{P})\) for the SAL model.

2) the optimal virtual battery model (OVB)~\cite{9979729}, an analytical inner-approximation method. For the STN model, we need to relax the decision variables to continuous variables as the OVB method only applies to linear constraints. Moreover, it cannot be utilized for steelmaking plants modeled by RTN due to the absence of linear relaxation methods.

\begin{table}[!t]\renewcommand{\arraystretch}{1}
  \caption{Performance Comparison of Different Methods}
  \label{tab_combined}\vspace{-2ex}
  \centering
  \setlength{\tabcolsep}{4pt}
  {\scriptsize
  \begin{tabular}{c|ccc|ccc}
    \toprule
    \multirow{2}{*}{Method} & \multicolumn{3}{c|}{Normalized RMSE (\%)} & \multicolumn{3}{c}{Continuous (Integer) Variables} \\
    \cmidrule{2-7}
    & Cement & \begin{tabular}[c]{@{}c@{}}Steel\\powder\end{tabular} & \begin{tabular}[c]{@{}c@{}}Steel-\\making\end{tabular} & Cement & \begin{tabular}[c]{@{}c@{}}Steel\\powder\end{tabular} & \begin{tabular}[c]{@{}c@{}}Steel-\\making\end{tabular} \\
    \midrule
    Original & - & - & - & 196(288) & 490(720) & 0(10208) \\
    OVB & 25.4 & 36.7 & N/A & 48(0) & 48(0) & N/A \\
    SAL & 10.2 & 23.0 & 8.3 & 24(0) & 24(0) & 24(0) \\
    D3R-1 & \textbf{8.9} & 17.2 & \textbf{3.6} & 24(0) & 24(0) & 24(0) \\
    D3R-2 & 9.5 & \textbf{10.3} & 4.2 & 48(0) & 48(0) & 48(0) \\
    \bottomrule
  \end{tabular}
  }
\end{table}

\begin{figure}[!t]
  \centering
  \includegraphics[width=3.0in]{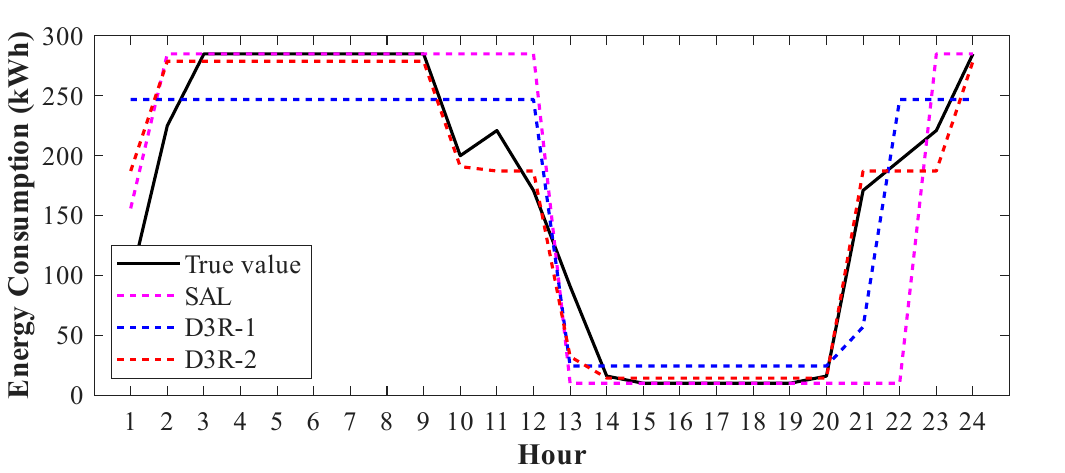}
  \vspace{-2ex}
  \caption{Optimal energy consumption results provided by the original model and the reduced constraints under the same electricity prices.}
  \label{fig_dispatch}
\end{figure}

\begin{figure}[!t]
  \centering
  \includegraphics[width=3.0in]{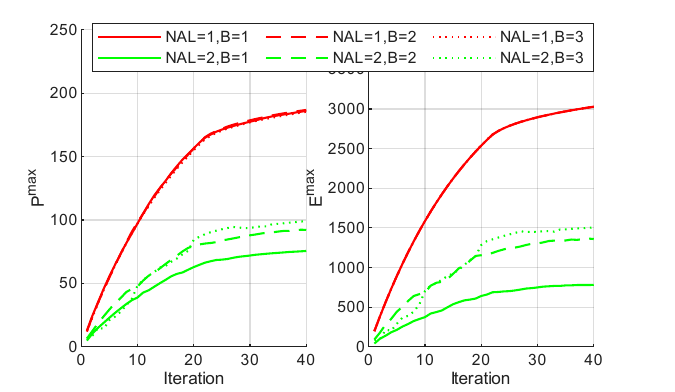}
  \includegraphics[width=3.0in]{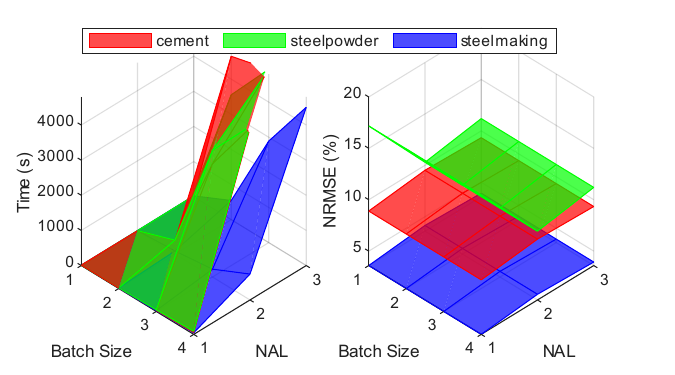}
  \vspace{-2ex}
  \caption{NAL: number of ALs. B: batch size. Top: Convergence of the ALF parameters during training for the steel powder plant case, exemplified by AL No.1. Bottom: Comparison of computation time and optimal energy consumption errors with different NAL and B.}
  \label{fig_convergence}
\end{figure}

\paragraph{Results and comparison} The proposed D3R achieved the best performance on all three datasets. This means that the dimensionality reduction constraints obtained through this method yielded the lowest normalized root mean square error (nRMSE) compared with the results based on the original constraints in the test set (Table~\ref{tab_combined}). Here, D3R-1(2) represents D3R via the ALF model with 1(2) ALs. The results from OVB based on the analytical inner approximation method were significantly inferior. Fig.~\ref{fig_dispatch} visually demonstrates the effectiveness of D3R, where it reduces the industrial users' complex constraints and variables in the optimization problem with an error ranging from 4\% to 10\% (Table~\ref{tab_combined}). Fig.~\ref{fig_convergence} illustrates the training process and computational efficiency of D3R. In the cement plant case (4 production stages), the OVB method converged after 767 seconds of iteration. However, in the steel powder plant case (10 production stages), the OVB did not converge in 24 hours; therefore, the values of the unconverted states were used.

\vspace{-2ex}
\section{Conclusion}\label{sec_conclusion}
Diverging from bottom-up analytical approaches, we propose D3R, a data-driven dimension reduction framework that trains a low-dimensional constraint model to fit the optimal solutions derived from the original constraints. D3R was tested on three industrial load datasets, and the results demonstrated that D3R reduced the high-dimensional constraints without losing much optimality, outperforming the analytical methods in terms of accuracy. D3R provides a new perspective for dimension reduction in industrial load modeling and other tasks with complex constraints to reduce, enhancing adaptability and compatibility with integer variables.

Future research could focus on: (1) developing more realistic forms of dimension-reduced constraints within the D3R framework to capture industrial load characteristics more accurately; (2) addressing how to decompose the overall energy consumption derived from reduced constraints into individual devices.


\bibliographystyle{IEEEtran}
\bibliography{reference}

@article{zhang_cost-effective_2017,
	title = {Cost-{Effective} {Scheduling} of {Steel} {Plants} {With} {Flexible} {EAFs}},
	volume = {8},
	issn = {1949-3053, 1949-3061},
	language = {en},
	number = {1},
	urldate = {2023-03-13},
	journal = {IEEE Trans. Smart Grid},
	author = {Zhang, Xiao and Hug, Gabriela and Harjunkoski, Iiro},
	month = jan,
	year = {2017},
	pages = {239--249},
}

@article{muller_aggregation_2019,
	title = {Aggregation and {Disaggregation} of {Energetic} {Flexibility} {From} {Distributed} {Energy} {Resources}},
	volume = {10},
	issn = {1949-3053, 1949-3061},
	doi = {10.1109/TSG.2017.2761439},
	language = {en},
	number = {2},
	urldate = {2023-04-12},
	journal = {IEEE Trans. Smart Grid},
	author = {Muller, Fabian L. and Szabo, Jacint and Sundstrom, Olle and Lygeros, John},
	month = mar,
	year = {2019},
	pages = {1205--1214},
}

@article{lu_data-driven_2021,
	title = {Data-driven real-time price-based demand response for industrial facilities energy management},
	volume = {283},
	issn = {0306-2619},
	language = {en},
	urldate = {2022-10-07},
	journal = {Appl. Energy},
	author = {Lu, Renzhi and Bai, Ruichang and Huang, Yuan and Li, Yuting and Jiang, Junhui and Ding, Yuemin},
	month = feb,
	year = {2021},
	pages = {116291},
}

@article{tan_non-iterative_2023,
	title = {Non-{Iterative} {Solution} for {Coordinated} {Optimal} {Dispatch} {Via} {Equivalent} {Projection}—{Part} {I}: {Theory}},
	issn = {1558-0679},
	shorttitle = {Non-{Iterative} {Solution} for {Coordinated} {Optimal} {Dispatch} {Via} {Equivalent} {Projection}—{Part} {I}},
	doi = {10.1109/TPWRS.2023.3258066},
	journal = {IEEE Trans. Power Syst.},
	author = {Tan, Zhenfei and Yan, Zheng and Zhong, Haiwang and Xia, Qing},
	year = {2023},
	pages = {1--9},
}

@misc{liu_primer_2020,
	title = {A {Primer} on {Zeroth}-{Order} {Optimization} in {Signal} {Processing} and {Machine} {Learning}},
	language = {en},
	urldate = {2024-08-21},
	publisher = {arXiv},
	author = {Liu, Sijia and Chen, Pin-Yu and Kailkhura, Bhavya and Zhang, Gaoyuan and Hero, Alfred and Varshney, Pramod K.},
	month = jun,
	year = {2020},
	note = {arXiv:2006.06224},
}

@misc{rick10119_data-driven_2024,
	author = {Ruike Lyu},
	url = {https://github.com/Rick10119/Data-Driven-Dimension-Reduction},
	urldate = {2024-09-07},
}

@inproceedings{lyu_approximating_2024,
	title = {Approximating {Energy}-{Regulation} {Feasible} {Region} of {Virtual} {Power} {Plants}: {A} {Data}-driven {Inverse} {Optimization} {Approach}},
	shorttitle = {Approximating {Energy}-{Regulation} {Feasible} {Region} of {Virtual} {Power} {Plants}},
	booktitle = {{2024 IEEE Power \& Energy Society General Meeting}},
	author = {Lyu, Ruike and Guo, Hongye and Chen, Qixin},
}

@article{golmohamadi_robust_2020,
	title = {Robust {Self}-{Scheduling} of {Operational} {Processes} for {Industrial} {Demand} {Response} {Aggregators}},
	volume = {67},
	language = {en},
	number = {2},
	journal = { IEEE Trans. Ind. Electron.},
	author = {Golmohamadi, Hessam and Keypour, Reza and Bak-Jensen, Birgitte and Pillai, Jayakrishnan R and Khooban, Mohammad Hassan},
	year = {2020},
}

@article{fortuny-amat_representation_1981,
	title = {A {Representation} and {Economic} {Interpretation} of a {Two}-{Level} {Programming} {Problem}},
	volume = {32},
	issn = {1476-9360},
	number = {9},
	journal = {J. Oper. Res. Soc},
	author = {Fortuny-Amat, José and McCarl, Bruce},
	month = sep,
	year = {1981},
	pages = {783--792},
}

@ARTICLE{9979729,
  author={Tan, Zhenfei and Yu, Ao and Zhong, Haiwang and Zhang, Xianfeng and Xia, Qing and Kang, Chongqing},
  journal={CSEE J. Power Energy Syst.}, 
  title={Optimal Virtual Battery Model for Aggregating Storage-Like Resources with Network Constraints}, 
  year={2024},
  volume={10},
  number={4},
  pages={1843-1847}}

@article{chen_leveraging_2021,
	title = {Leveraging {Two}-{Stage} {Adaptive} {Robust} {Optimization} for {Power} {Flexibility} {Aggregation}},
	volume = {12},
	issn = {1949-3053, 1949-3061},
	doi = {10.1109/TSG.2021.3068341},
	language = {en},
	number = {5},
	urldate = {2023-04-12},
	journal = {IEEE Trans. Smart Grid},
	author = {Chen, Xin and Li, Na},
	month = sep,
	year = {2021},
	pages = {3954--3965},
}

@article{chen_aggregate_2020,
	title = {Aggregate {Power} {Flexibility} in {Unbalanced} {Distribution} {Systems}},
	volume = {11},
	issn = {1949-3061},
	doi = {10.1109/TSG.2019.2920991},
	number = {1},
	urldate = {2023-11-06},
	journal = {IEEE Trans. Smart Grid},
	author = {Chen, Xin and Dall'Anese, Emiliano and Zhao, Changhong and Li, Na},
	month = jan,
	year = {2020},
	pages = {258--269},
}

\end{document}